\documentclass[journal]{IEEEtran}
\ifCLASSINFOpdf
\else
\fi
\usepackage{booktabs}
\usepackage{array}    
\usepackage{amsmath,amssymb,amsfonts}
\usepackage{graphicx}
\usepackage{textcomp}
\usepackage{cite}
\usepackage{xcolor}
\usepackage{afterpage}
\usepackage{multirow}
\usepackage{bm}
\usepackage{tabularx} 
\usepackage{stfloats} 
\usepackage{soul}
\usepackage{enumitem}
\usepackage{hyperref}
\usepackage{fontawesome5}
\usepackage{hhline}
\usepackage{siunitx}
\usepackage{caption}

\usepackage{varwidth} 
\usepackage{ragged2e} 
\usepackage{booktabs}   

\usepackage{algorithm}
\usepackage{algpseudocode}
\usepackage{algorithmicx}
\usepackage{amsmath}
\usepackage{amssymb}
\usepackage{calc}  

\algrenewcommand\algorithmicrequire{\textbf{Require:}}
\algrenewcommand\algorithmicrepeat{\textbf{repeat}}
\algrenewcommand\algorithmicuntil{\textbf{until}}

\newlength{\RequireLabelWidth}
\newcommand{\RequireH}[1]{%
  \Statex\hspace*{-\algorithmicindent}
  \begingroup
    \settowidth{\RequireLabelWidth}{\textbf{Require: }}%
    \hangafter=1\hangindent=\RequireLabelWidth
    \parindent=0pt
    \noindent\textbf{Require: }#1\par
  \endgroup
}
\newcommand{\InputH}[1]{%
  \Statex\hspace*{-\algorithmicindent}%
  \begingroup
    \settowidth{\RequireLabelWidth}{\textbf{Input: }}%
    \hangafter=1\hangindent=\RequireLabelWidth
    \parindent=0pt
    \noindent\textbf{Input: }#1\par
  \endgroup
}
\newcommand{\OutputH}[1]{%
  \Statex\hspace*{-\algorithmicindent}%
  \begingroup
    \settowidth{\RequireLabelWidth}{\textbf{Output: }}%
    \hangafter=1\hangindent=\RequireLabelWidth
    \parindent=0pt
    \noindent\textbf{Output: }#1\par
  \endgroup
}
\newcommand{\StateIndent}[1]{%
  \State\hspace{1em}\parbox[t]{\dimexpr\linewidth-1.2em\relax}{#1}%
}

\def\BibTeX{{\rm B\kern-.05em{\sc i\kern-.025em b}\kern-.08em
    T\kern-.1667em\lower.7ex\hbox{E}\kern-.125emX}}

\begin{document}
%
\title{From Similarity to Feasibility: Diffusion-Refined Retrieval-Augmented Generation for Distribution Network Optimization}
%
%
%


\author{Yuxuan Chen, Haipeng Xie, \IEEEmembership{Senior Member, IEEE}, Shuo Dai, Ruoyi Xu, Zhaohong Bie, \IEEEmembership{Fellow, IEEE}
\thanks{All authors are with the National Key Laboratory for High Energy Pulsed Power, Xi’an Jiaotong University, Xi’an, Shaanxi, China. (corresponding author: H. Xie; e-mail: haipengxie@xjtu.edu.cn)}}

%
%

\markboth{IEEE/CAA JOURNAL OF AUTOMATICA SINICA,~Vol.~X, No.~X, X~X}%
{Xxx \MakeLowercase{\textit{et al.}}: From Similarity to Feasibility: Diffusion-Refined Retrieval-Augmented Generation for Distribution Network Optimization}
%



\maketitle

\begin{abstract}
Rapidly shifting operational scenarios driven by uncertain Distributed Energy Resource (DER) profiles render conventional distribution network optimization methods either computationally expensive or poorly generalizable. This paper introduces GridRAG, a pioneering retrieval-augmented framework that transforms optimization into a ``retrieve-and-refine'' paradigm. GridRAG first embeds scenario features and optimal solutions into a joint representation space to ensure semantic consistency. Based on the hybrid semantic information, the similar historical scenarios are then retrieved from a pre-constructed database. Then an SDEdit-style diffusion module is integrated to refine retrieved solutions by modeling the conditional distribution over near-feasible manifolds. This process effectively pulls retrieved solutions into near-optimal attraction basins, providing a high-quality warm-start for the final solver. Validated on three optimization tasks across four standard topologies, GridRAG demonstrates superior cross-scenario generalization and a multi-fold speedup in solution time compared to existing learning-based and model-based baselines. Our code is available at https://anonymous.4open.science/r/GridRAG-A328.
\end{abstract}


%
\IEEEpeerreviewmaketitle

\section{Introduction}
%
%
%
%
\IEEEPARstart{T}{he} rapid integration of DERs has led to a ``scenario explosion'' in modern distribution networks~\cite{joshi2023survey}. This trend has fundamentally reshaped a wide range of optimization tasks, including volt/var control (VVC/VVO), optimal power flow, network reconfiguration, coordinated active-reactive power optimization, etc~\cite{hossain2023machine}. Consequently, there is an urgent demand for optimization frameworks that possess both high computational efficiency and robust generalization capabilities.

Existing methods for distribution network optimization can be broadly categorized into two groups. \textbf{Model-based} approaches formulate optimization problems through detailed physical modeling and solve them using commercial solvers~\cite{capitanescu2007contingency,li2008decomposed,weinhold2020fast,tejada2017security}. While highly interpretable, these methods are often computationally expensive and tailored to specific scenarios. \textbf{AI-based} approaches have gained traction by leveraging Deep Neural Networks (DNNs)~\cite{pan2020deepopf,lei2020data,pineda2020data,huang2022applications} or Reinforcement Learning (RL)~\cite{yang2019two,zhang2020deep,hu2022multi,chen2025robust} to predict active constraints or directly map scenarios to solutions. While data-driven methods offer fast online inference, they typically suffer from a ``one model per scenario'' limitation, requiring extensive retraining for novel conditions. Although recent advancements in transfer learning, meta-learning, and hybrid data-knowledge-driven strategies have improved cross-scenario performance~\cite{liu2022varying,gao2023data,li2021meta,xiao2022meta,feng2024short,ding2024load,luo2023generalizable}, they remain largely effective only for in-distribution adaptation and struggle to generalize to severe Out-of-Distribution (OOD) scenarios (zero-shot).

\begin{figure}[!t]
\centering
\includegraphics[width=\columnwidth]{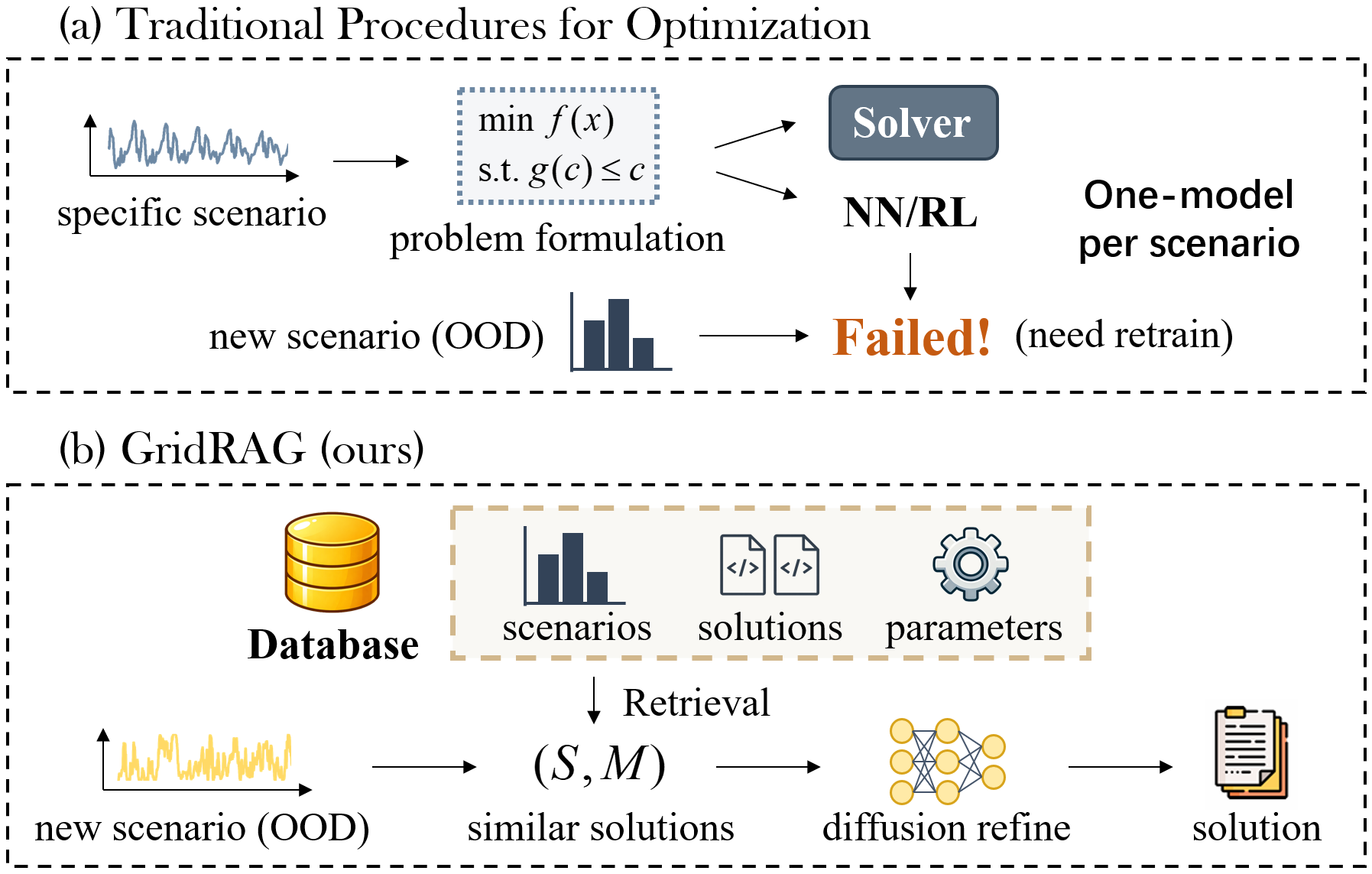}
\caption{(a) Traditional procedures require a customized model for each specific scenario; (b) Our method leverages retrieval augmentation to address scenario explosion in distribution network optimization.}
\label{fig:1}
\vspace{-1.0em}
\end{figure}

Retrieval-Augmented Generation (RAG) is an emerging paradigm widely used in large language models to guide high-quality generation by retrieving task-relevant information from massive databases~\cite{lewis2020retrieval}. Recent advances in text-to-motion and image generation have further demonstrated the potential of retrieval techniques in enhancing generative efficiency and quality~\cite{zhang2023remodiffuse,chenre}. These successes motivate us to explore retrieval augmentation for distribution network optimization. However, directly migrating RAG to power system optimization faces two significant challenges: \textbf{Firstly}, unlike natural language, operating scenarios and optimization solutions represent distinct modalities with non-bijective mappings; thus, similarity must reflect decision transferability rather than mere scenario feature proximity. \textbf{Secondly}, unlike web-scale corpora, engineering databases are inherently finite. Under high DER stochasticity, OOD scenarios can weaken retrieval effectiveness and even induce negative transfer, necessitating a robust zero-shot refinement mechanism.

In this paper, we revisit the RAG-based optimization problem from a distinct perspective of multi-modal representation learning. In our vision, the crux of effective retrieval is not simply scenario similarity, but lies in the alignment of scenario-solution representations in a shared multi-modal space, which naturally preserves the underlying decision semantics. Furthermore, a key finding from our multi-modal experiments reveals that even when a scenario modality shifts significantly, the retrieved solutions may still remain within the ``near-feasible'' manifold. This observation leads to a central hypothesis: \textit{the key to zero-shot optimization is not predicting the unique optimum directly, but rapidly entering the attraction basin of feasible and near-optimal manifolds}, which require us to model the distribution of promising solutions. Generative models are particularly suitable for this purpose, and among them, diffusion models have recently demonstrated superior performance~\cite{chen2024overview}. We therefore utilize diffusion model to learn a conditional distribution, where retrieved solutions are refined into a probabilistic warm-start, ensuring both rapid convergence and physical feasibility.

To model these insights, we propose GridRAG, a novel framework for distribution network optimization (Fig. 1). Unlike traditional optimization pipelines that require per-scenario customization, GridRAG fully exploits retrieval augmentation to address scenario explosion in distribution network optimization. Its architecture is built upon three key innovations: 1) \textbf{Grid2Vec} embeds scenarios and solutions into a joint representation space, capturing fine-grained semantic correlations. 2) \textbf{Hybrid Retrieval} identifies feasible solution-oriented retrieval, progressing from \textit{Fast Recall} to \textit{Precise Reranking} based on the learned joint embeddings. 3) \textbf{Similarity-Gated Adaptation} controls refinement intensity according to retrieval similarity. An SDEdit-style diffusion procedure treats the retrieved solution as an initial state and applies adaptive diffusion with iterative refinement, forming an optimal attraction path toward a warm-start feasible region.

Our contributions are as follows:

1) We propose GridRAG, the first retrieval-augmented framework for distribution network optimization. By integrating multi-modal representation learning, retrieval, and adaptive refinement, GridRAG enables fast and accurate solution generation across diverse and OOD scenarios.

2) We further introduce an SDEdit-style diffusion refinement module that treats retrieved solutions as initial states, significantly enhancing the robustness of RAG-optimization under zero-shot settings.

3) We evaluate GridRAG on a comprehensive, large-scale benchmark encompassing diverse tasks and varying distribution network topologies. To the best of our knowledge, this represents the most extensive evaluation benchmark in the field of learning-based power system optimization to date. Results establish that GridRAG consistently outperforms strong baselines in both solution accuracy and time efficiency.

\section{Related Works}
\subsection{Traditional Approaches for Optimization}
Distribution network optimization is typically formulated as mixed-integer nonlinear programs under AC-OPF/DistFlow constraints, which can be broadly classified into two families by solution methodology. Model-based approaches construct detailed optimization models and solve them using solvers. Research in this area has largely focused on computational acceleration, e.g., convex relaxations or linearized approximations~\cite{capitanescu2007contingency}, decomposition and distributed solution techniques such as Benders and ADMM~\cite{li2008decomposed,weinhold2020fast}, and heuristic pre-screening rules to reduce the search space~\cite{tejada2017security}. Model-based methods offer strong interpretability and feasibility guarantees, but are typically confined to single scenarios and suffer from long online solve times.

Data-driven approaches fall into two main paradigms: supervised DNN mappings and reinforcement learning (RL). DNN mappings either predict full candidate~\cite{pan2020deepopf,lei2020data} or predict likely active constraint sets to shrink the optimization~\cite{pineda2020data,huang2022applications}. Having the advantage of handling large state-action spaces, deep RL has been widely applied to complex distribution control tasks. DQN updates Q-values via DNNs, and multi-agent DQN has been shown to be effective for VVC/VVO tasks in distribution networks~\cite{yang2019two,zhang2020deep}. Policy-based and actor-critic methods (e.g., DDPG, PPO) are used for hybrid or continuous controls. Many studies apply these algorithms to reactive power control and VVO problems~\cite{hu2022multi,chen2025robust}. Overall, data-driven methods fit nonlinear mappings and enable fast online inference, but they lack interpretability and can violate physical constraints. Beyond the above, neither model-based nor standard data-driven approaches provide reliable cross-scenario generalization: they typically require re-tuning or retraining, which is a critical weakness under the explosion of operating scenarios.

\subsection{Cross-Scenario Generalization Strategies}
Several studies focus on leveraging historical knowledge to accelerate optimization in novel scenarios. One prominent research thrust employs hybrid data-knowledge-driven paradigms to automate knowledge extraction. For instance, Liu, et al.~\cite{liu2022varying} utilize DNNs to predict critical security constraints while managing diverse operating conditions and their corresponding models via Knowledge Graph (KG). Similarly, Gao, et al.~\cite{gao2023data} employ KG to archive model parameters across historical scenarios, determining whether to reuse, fine-tune, or retrain the model based on the similarity between new and historical scenarios. While these approaches effectively capitalize on historical experience, they suffer from limited flexibility due to manually defined similarity thresholds. Furthermore, they typically necessitate model retraining when encountering unprecedented scenarios.

Several meta-learning and transfer-learning based studies report comparable transferability. Li, et al.~\cite{li2021meta} adopt a meta-learning strategy that trains an offline classifier to automatically select an appropriate model for a new task, enabling online load forecasting; however, their framework relies on a manually specified Pearson-correlation threshold, which restricts adaptability. Xiao, et al.~\cite{xiao2022meta} and Feng, et al.~\cite{feng2024short} leverage meta-learning to speed up online adaptation, yet their approaches fundamentally rely on local fine-tuning and thus do not support true zero-shot transfer. The works of Ding, et al.~\cite{ding2024load} and Luo, et al.~\cite{luo2023generalizable} use similar strategies as retrieval: they construct a support set in the source domain and perform prototype-based similarity matching for target-domain inference, avoiding target-side fine-tuning for non-intrusive load monitoring. Nevertheless, these meta-learning methods primarily reuse prior knowledge and offer limited genuine zero-shot generalization. Consequently, they struggle to guarantee robust inference under severe out-of-distribution shifts at test time.

\section{Problem Statement}

In this study, we aim to develop an end-to-end learning system $\mathcal{F}(\cdot)$ for solving a variety of optimal operation problems in distribution networks.

Given a new operating scenario ${{S}_{\text{new}}}$ and a pre-constructed database $\mathcal{D}$ of historical scenario-solution pairs $(S,\bm{M})$, the framework generates a set of predicted control decisions:

\vspace{-0.5em}
\begin{equation}\label{eq:1}
\hat{\mathbb{M}}=\mathcal{F}({{S}_{\text{new}}},\mathcal{D})=\{{{\hat{\bm{m}}}_{1}},...,{{\hat{\bm{m}}}_{L}}\}\in {\mathbb{R}^{C\times L}}
\end{equation}

\noindent
where $\hat{\mathbb{M}}$ represents the predicted solution for the target scenario, and each ${{m}_{l}}$ at time step $l$ is a control action vector comprising continuous control variables $\bm{u}_{l}^{\text{cont}}\in {{\mathbb{R}}^{{{C}_{\text{cont}}}}}$ and discrete control variables $\bm{u}_{l}^{\text{disc}}\in {{\mathbb{Z}}^{{{C}_{\text{disc}}}}}$.

The continuous variables $u_{l}^{\text{cont}}$ typically include energy storage charge/discharge power ${{P}_{\text{ess}}}$, reactive power from static var compensators (SVC) ${{Q}_{\text{svc}}}$, etc. The discrete variables $u_{l}^{\text{disc}}$ include OLTC tap positions ${{\delta }_{\text{oltc}}}$, capacitor bank steps ${{Q}_{\text{sc}}}$, and other switching actions, with the exact composition varying across tasks.

It is worth noting that the primary focus of this study is to assess the generalization performance of learned models rather than developing a specialized solver tailored to any single operating condition. After offline training, the proposed method is evaluated on previously unseen scenarios, where we compare it against strong baseline approaches in terms of time consumption, solution quality, constraint satisfaction, and other key performance metrics.

\begin{figure*}[!t]
\centering
\includegraphics[width=\textwidth]{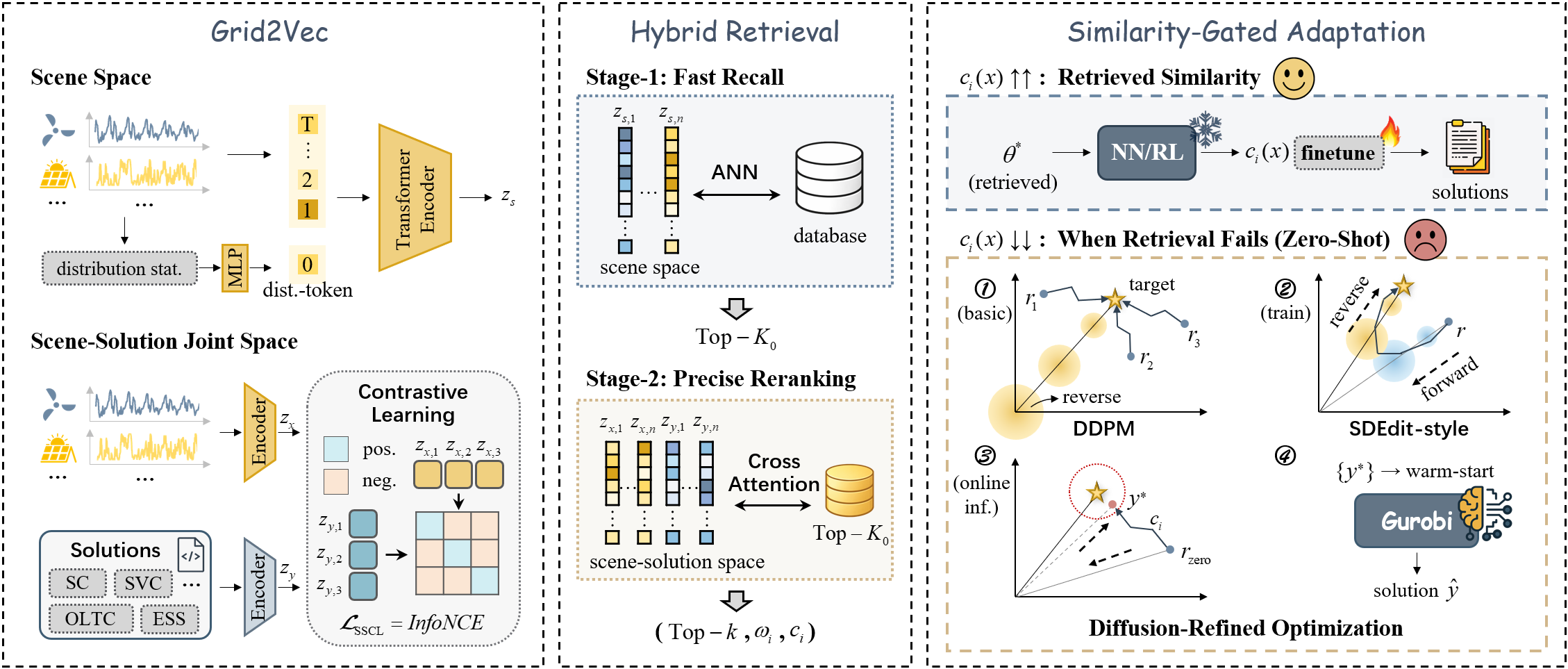}
\caption{The overall framework of GridRAG.}
\label{fig:2}
\end{figure*}

\section{Methodology}
The overall architecture of the proposed GridRAG framework is illustrated in Fig. 2. GridRAG leverages retrieval augmentation to learn from historical scenarios, enabling fast cross-scenario generalization through lightweight adaptation and diffusion-based refinement. The framework consists of three main components:

\begin{itemize}[leftmargin=9pt]
  \item \textbf{Grid2Vec.} This module embeds complex distribution network operating scenarios into a latent space, capturing comprehensive and fine-grained feature semantic representations. Contrastive learning is employed to enhance scenario-solution consistency in the joint embedding space.
  \item \textbf{Hybrid Retrieval.} We adopt a two-stage retrieval paradigm for efficiency. \textit{Fast Recall} performs coarse retrieval in the scenario feature space. And \textit{Precise Reranking} operates in the joint embedding space using attention-based fusion to retrieve top-k similar cases and produce a retrieval confidence score, which softly conditions downstream modules.
  \item \textbf{Similarity-Gated Adaptation.} Execution of the downstream task is gated by the retrieval confidence. This parameter will control the intensity of fine-tuning and diffusion refinement. We design an SDEdit-style diffusion module which treats the retrieved solution as initial stage and pulls it toward the feasible neighborhood, enabling a better warm-start for the downstream optimizer.
\end{itemize}

\vspace{-0.4em}
\subsection{Grid2Vec}
We formulate the retrieval-augmented optimization as a multi-modal learning task. We argue that standard pattern recognition in the raw scenario modality does not necessarily map to the scenario-solution manifold. To preserve authentic motion semantics, we define \textit{scenario similarity} not by input features alone, but by the similarity of their optimal solutions, thereby ensuring decision transferability. Consequently, constructing a robust embedding space is critical for the downstream pipeline.

Grid2Vec consists of two parallel channels: 1) a \textbf{Scene-only} encoder that learns retrieval-ready representations from DER temporal dynamics and statistical distributions for large-scale fast recall; 2) a \textbf{Scene-Solution} aligned encoder that aligns scenarios and solutions via contrastive learning.

\vspace{0.3em}
\noindent
\textbf{Scene Space Representation.} While modeling temporal dynamics is necessary, downstream optimization is particularly sensitive to distribution shifts. Therefore, we adopt an architecture combining temporal and distributional features:

\begin{itemize}[leftmargin=9pt]
  \item \textbf{Temporal Branch:} A 1D-CNN is employed to extract temporal tokens, denoted as ${{h}_{\text{seq}}}$.
  \item \textbf{Distribution Branch:} For each channel, we extract a set of statistical features, including moments (mean, variance, skewness), quantiles, and peak features. These statistics are concatenated and processed by a MLP layer to generate distribution tokens ${{h}_{\text{dist}}}$.
\end{itemize}

Finally, a Transformer encoder fuses these tokens to produce the scenario embedding:

\vspace{-0.5em}
\begin{equation}\label{eq:1}
{{z}_{s}}={{E}_{s}}({{h}_{\text{seq}}},{{h}_{\text{dist}}})
\end{equation}

\noindent
\textbf{Scene-Solution Joint Space.} We employ lightweight Transformers as encoders to embed both scenario and solution (time × variable matrix) into a unified joint space:

\vspace{-0.5em}
\begin{equation}\label{eq:1}
{{z}_{x}}={{E}_{x}}(x)\text{ },\text{ }{{z}_{y}}={{E}_{y}}(y)
\end{equation}

To align these modalities, we apply CLIP-style contrastive training using InfoNCE loss. Within a batch, pairs $(z_{x}^{i},z_{y}^{i})$ are treated as positive samples, while $(z_{x}^{i},z_{y}^{j})$ (where $i \neq j$) serve as negative samples. The training objective maximizes the similarity score $\left\langle z_{x}^{i},z_{y}^{i} \right\rangle$ relative to the negative pairs $\left\langle z_{x}^{i},z_{y}^{j} \right\rangle$.

\vspace{-0.3em}
\subsection{Hybrid Retrieval}
The efficacy of retrieval augmentation hinges on identifying relevant and actionable historical information. To this end, we design a two-stage hybrid retrieval framework that transitions from large-scale \textit{Fast Recall} to \textit{Precise Reranking}, ultimately outputting a continuous retrieval confidence score. This approach enables end-to-end automated knowledge acquisition without relying on manually predefined thresholds.

\vspace{0.3em}
\noindent
\textbf{Fast Recall.} Our primary objective in this stage is to rapidly narrow the search space from the entire database $\mathcal{D}$ to a candidate subset of size $\text{Top}-{{K}_{0}}$. We perform Approximate Nearest Neighbor (ANN) search using the scene-only embeddings. For efficient engineering implementation, the Hierarchical Navigable Small World (HNSW) algorithm is employed. This phase prioritizes high recall over absolute precision to ensure that potentially optimal candidates are not prematurely discarded.

\vspace{0.3em}
\noindent
\textbf{Precise Reranking.} we refine the candidate set to produce a final similarity ranking and a confidence metric. We design a cross-attention module, as illustrated in Fig. 3, which operates in the scene-solution joint space. The input scenario embedding ${{z}_{x}}$ serves as the Query, while the embeddings of the $\text{Top}-{{K}_{0}}$ candidates act as Keys and Values. We utilize the resulting attention weights as the reranking scores ${{s}_{\text{score}}}$, from which the most similar $\text{Top}-k$ scenarios are selected.

To govern the downstream adaptation intensity, we introduce a confidence network $g_c$ consisting of a two-layer MLP that outputs the retrieval confidence $c$:

\vspace{-0.5em}
\begin{equation}\label{eq:1}
c({{z}_{x}})=\sigma ({{g}_{c}}([||\delta ||,s_{\text{score}}^{1},\Delta s,\bar{s}_{\text{score}}^{1:k}]))
\end{equation}

\noindent
where $\delta$ denotes the distance between the input scenario and the cluster centroid, $s_{\text{score}}^{1}$ is the top-1 similarity score, $\Delta s=s_{\text{score}}^{1}-s_{\text{score}}^{2}$ represents the margin between the first and second candidates, and $\bar{s}_{\text{score}}^{1:k}$ is the average score of the top-k candidates.

\begin{figure}[!t]
\centering
\includegraphics[width=0.95\columnwidth]{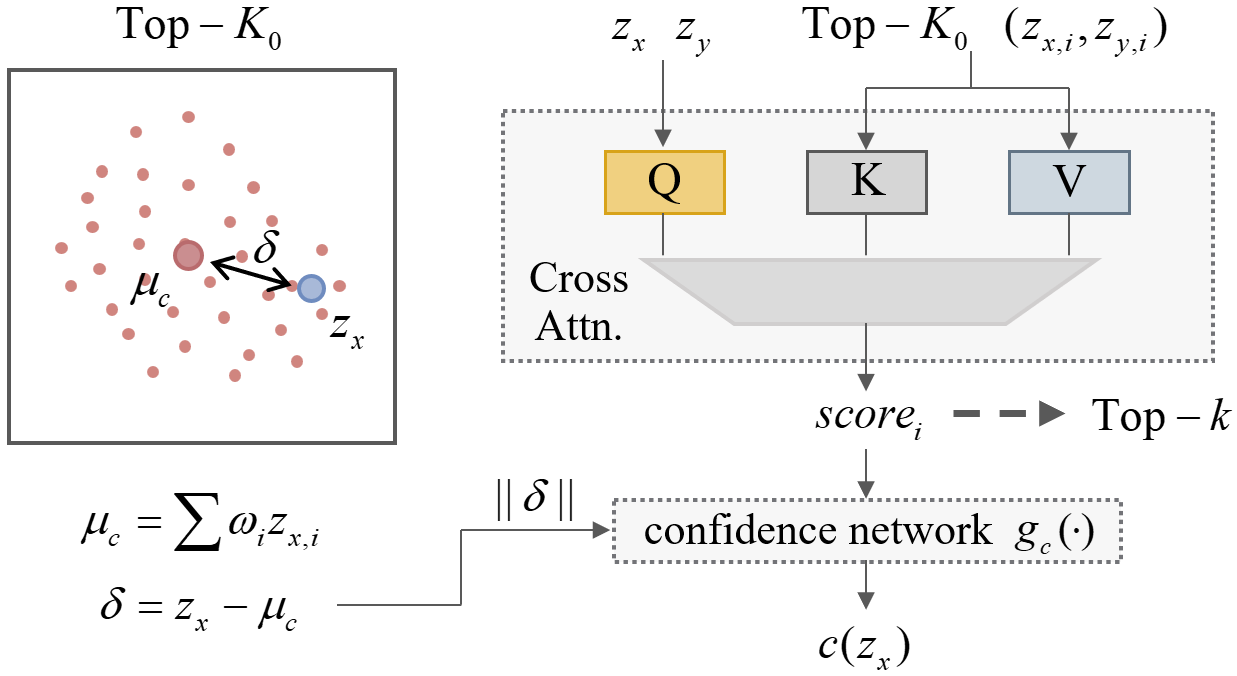}
\caption{Architecture of precise reranking module}
\label{fig:3}
\vspace{-0.6em}
\end{figure}

By parameterizing the confidence estimation, we move beyond traditional heuristic thresholding~\cite{gao2023data}. This allows $c$ to function as a continuous controller for downstream execution, which will be trained from end-to-end throughout the pipeline.

\vspace{0.3em}
\subsection{Similarity-Gated Adaptation}
Once relevant historical scenarios are retrieved, the framework must effectively leverage this information for downstream optimization. We employ a confidence-gated mechanism to control the refinement intensity: for high-confidence retrievals (within clusters), a lightweight adaptation is applied following~\cite{liu2022varying,gao2023data}. Our primary contribution, however, lies in addressing the zero-shot challenge——where the retrieved scenario falls outside any historical clusters.

\vspace{0.3em}
\noindent
\textbf{Motivation: From Near-Feasible Priors to Optimal Basins}


Our multi-modal analysis reveals that even in zero-shot cases, the retrieved solution may still reside in a ``near-feasible'' manifold of the target scenario. This phenomenon motivates us to rethink zero-shot generalization: the key is not to predict a unique optimal solution directly but to rapidly enter the attraction basin of the near-optimal manifold.

So this task indicates a conditional distribution modeling of $p(\bm{M}|S,c)$, where diffusion model excels by iteratively refining noise into structured data. By treating the refined solution as a probabilistic warm-start, our diffusion-refined optimization transforms the retrieval result into a high-quality initial point for the final solver.

\vspace{0.3em}
\noindent
\textbf{Diffusion Backbone and Physical Guidance}

We first train a diffusion backbone to model the solution distribution. Let ${\bm{M}^{0}}$ denote the clean solution and ${\bm{M}^{t}}$ the noisy state at diffusion step $t$. Following the DDPM paradigm~\cite{ho2020denoising}, the forward process $q$ can be expressed as:

\vspace{-0.5em}
\begin{equation}\label{eq:1}
q({\bm{M}^{t}}|{\bm{M}^{t-1}}):=N({\bm{M}^{t}};\sqrt{1-{{\beta }_{t}}}{\bm{M}^{t-1}},{{\beta }_{t}}\bm{I})
\end{equation}

\vspace{-0.5em}
\begin{equation}\label{eq:1}
q({\bm{M}^{t}}|{\bm{M}^{0}}):=N({\bm{M}^{t}};\sqrt{{{{\bar{\alpha }}}_{t}}}{\bm{M}^{0}},(1-{{\bar{\alpha }}_{t}})\bm{I})
\end{equation}

\noindent
where ${{\bar{\alpha }}_{t}}=\prod\nolimits_{s=1}^{t}{{{\alpha }_{s}}}$, ${{\beta }_{t}}\in [0,1]$ and ${{\alpha }_{t}}=1-{{\beta }_{t}}$. So, ${\bm{M}^{t}}$ can be obtained from ${\bm{M}^{0}}$:

\vspace{-0.5em}
\begin{equation}\label{eq:1}
{\bm{M}^{t}}=\sqrt{{{{\bar{\alpha }}}_{t}}}{\bm{M}^{0}}+\sqrt{1-{{{\bar{\alpha }}}_{t}}}\varepsilon \text{ ,}\varepsilon \in N(0,\bm{I})
\end{equation}

The reverse process $p_\theta$ is still a Markov process and is defined as:

\vspace{-0.5em}
\begin{equation}\label{eq:1}
{{p}_{\theta }}({\bm{M}^{t-1}}|{\bm{M}^{t}})=N({\bm{M}^{t-1}};{{\mu }_{\theta }}({\bm{M}^{t}},t),{{\Sigma }_{\theta }}({\bm{M}^{t}},t))
\end{equation}

\noindent
where ${{\Sigma }_{\theta }}$ is fixed as $\sigma _{t}^{2}I$. The denoising network ${{\mu }_{\theta }}$ is implemented as a Multi-head Transformer with sinusoidal position encoding to handle the sequential nature of the control variables~\cite{zhang2023remodiffuse}.

To ensure the generated solutions respect grid physics, we incorporate Energy-Guided Denoising~\cite{huang2023diffusion}. We augment the reverse transition with a guidance term, representing the gradient of physical and task-specific constraints:

\vspace{-0.5em}
\begin{equation}\label{eq:1}
{{p}_{\theta }}({\bm{M}^{t-1}}|{\bm{M}^{t}},S)=N({\bm{M}^{t-1}};{{\mu }_{\theta }}({\bm{M}^{t}},t)+\lambda \Sigma \bm{g},{{\Sigma }_{\theta }}({\bm{M}^{t}},t))
\end{equation}

\noindent
where $\lambda $ is a scaling factor of the guidance, $\bm{g}$ represents the Taylor expansion of the diffusion target ${{p}_{\phi }}({\bm{M}^{0}}|{\bm{M}^{t}},S)$ around ${\bm{M}^{t}}={{\mu }_{\theta }}$:

\vspace{-0.5em}
\begin{equation}\label{eq:1}
\log {{p}_{\phi }}({\bm{M}^{0}}|{\bm{M}^{t}},S)\approx ({\bm{M}^{t}}-{{\mu }_{\theta }})g+C
\end{equation}

\vspace{-0.5em}
\begin{equation}\label{eq:1}
\bm{g}={{\nabla }_{{\bm{M}^{t}}}}({{\varphi }_{g}}({\bm{M}^{t}}|S)+{{\varphi }_{p}}({\bm{M}^{t}}|S,\mathcal{O})){{|}_{{\bm{M}^{t}}={{\mu }_{\theta }}}}
\end{equation}

\noindent
where ${{\varphi }_{g}}$ is the objective for optimizing the solution with the specific scene, which is independent from the distribution network optimization target $\mathcal{O}$. It typically encompasses potential physical constraints (such as power flow constraints, operational limits, etc.)

The training objective of the diffusion backbone is to minimize the guided epsilon-error:

\vspace{-0.5em}
\begin{equation}\label{eq:1}
\mathcal{L}(\varepsilon )={{\mathbb{E}}_{{\bm{M}^{0}},t,\varepsilon }}[||\varepsilon -{{\varepsilon }_{\theta }}({\bm{M}^{t}},t,S)-\Sigma \bm{g}|{{|}^{2}}]
\end{equation}

\vspace{0.3em}
\noindent
\textbf{RAG-SDEdit for Adaptive Refinement}

In RAG paradigm, we aim to train the model to pull retrieved solutions ${\bm{M}_{\text{r}}}$ toward the ground truth $(S,{\bm{M}^{0}})$ using an SDEdit-style approach. The core of SDEdit-style is to use the retrieval solution as the initial state. In this stage, we train the reverse process by minimizing the MSE loss of $({{\hat{\bm{M}}}^{0}},{\bm{M}^{0}})$. Crucially, the diffusion depth $t$ is dynamically coupled with the retrieval confidence $c$:

\vspace{-0.5em}
\begin{equation}\label{eq:2}
t\sim \text{Uniform}(0,{{t}}(c)),\text{ }{{t}}(c)={{t}_{\min }}+(1-c)({{t}_{\max }}-{{t}_{\min }})
\end{equation}

A lower confidence $c$ assigns a larger $t$, granting the model more ``degrees of freedom'' to restructure the solution; conversely, a higher $c$ preserves the retrieved structure with minimal perturbations. During training, the confidence score $c$ is encoded via a two-layer MLP $\text{En}{{\text{c}}^{c}}$ and injected into the denoising network.

At test time, given a new scenario ${{S}_{\text{new}}}$, we retrieve ${\bm{M}_{\text{r}}}$ and execute the forward-reverse chain. The resulting ${{\hat{\bm{M}}}^{0}}$ serves as a high-fidelity warm-start for the downstream solver, ensuring rapid convergence to the global optimum. The complete procedures are summarized in Algorithm 1 (Training) and Algorithm 2 (Inference).

\begin{algorithm}
\caption{{Training of RAG-SDEdit Diffusion}}
\begin{algorithmic}[1]
\RequireH{Database $\mathcal{D}$; Number of diffusion steps ${N}$.}

\State Sample $(S,\bm{M}_0)$ from $\mathcal{D}$;
\State Sample $t\sim \text{Uniform}(0,{{t}}(c))$;
\State ${\bm{M}_{\text{r}}},c\leftarrow \text{Retrieval}(\mathcal{D};S,{\bm{M}^{0}})$;
\State $\bm{M}_{\text{r}}^{t}\leftarrow \text{Forward}({\bm{M}_{\text{r}}},t;c)$;
\State ${{\hat{\bm{M}}}^{0}}\leftarrow \text{Denoise}(\bm{M}_{\text{r}}^{t},S,t;c)$;
\State Update $\text{Denoise}(\cdot )$: $\nabla \mathcal{L}({{\hat{M}}^{0}},{{M}^{0}})$

\end{algorithmic}
\end{algorithm}
\vspace{-0.3em}

\begin{algorithm}
\caption{{Inference of RAG-SDEdit Diffusion}}
\begin{algorithmic}[1]
\InputH{New Scene $S_\text{new}$; Database $\mathcal{D}$; Steps ${N}$.}
\OutputH{${{\hat{\bm{M}}}^{0}}$}

\State ${\bm{M}_{\text{r}}},c\leftarrow \text{Retrieval}(\mathcal{D};S_\text{new})$;
\State $\bm{M}_{\text{r}}^{t}\leftarrow \text{Forward}({\bm{M}_{\text{r}}},t;c)$;
\StateIndent {\textbf{while} $t>0$ \textbf{do}:}
\StateIndent{\hspace{0.5em} ${{\hat{\bm{M}}}^{0}}\leftarrow \text{Denoise}(\bm{M}_{\text{r}}^{t},S_\text{new},t;c)$}
\StateIndent{\hspace{0.5em} $t\leftarrow t-\frac{1}{N}$}
\StateIndent{\hspace{0.5em} $\bm{M}_{\text{r}}^{t}\leftarrow \text{Forward}({\hat{\bm{M}}}^{0},t;c)$}
\StateIndent {\textbf{end}}
\State \textbf{return} ${{\hat{\bm{M}}}^{0}}$

\end{algorithmic}
\end{algorithm}
\vspace{-0.3em}

\section{Experiments}

\subsection{Experiment Setup}

\noindent
\textbf{Tasks.} To validate the universality of GridRAG, we design three distinct operational tasks spanning four standard topologies (IEEE-13, 33, 69, and 123 bus systems). Task A (Volt-Var Control): A mixed-integer problem minimizing power loss, considering discrete controllable devices like OLTCs and continuous ones like SVCs. Task B (Economic Dispatch): A continuous NLP problem minimizing operational costs while strictly adhering to network security constraints and OPF limits. Task C (Coordinated Active-Reactive Power Optimization): A highly complex task incorporating flexible resources to minimize a mixed operational cost.

\vspace{0.3em}
\noindent
\textbf{Datasets and Database.} We utilize a publicly available distribution network dataset DDRE-33~\cite{chen2025large}, which provides diverse DER generation profiles. DDRE-33 provides interpretable labels reflecting renewable characteristics that facilitate clear scenario clustering. This allows us to systematically design two test settings: \textbf{Test Group 1:} Target scenarios fall within established database clusters. \textbf{Test Group 2 (Zero-Shot):} Target scenarios correspond to label combinations strictly excluded from the training phase, falling entirely outside existing database clusters.

We allocate 80\% of the unique label combinations and split at a 9:1 ratio to form the database and Test Group 1. Each label combination has 2000 scenarios. The remaining 20\% are reserved for Test Group 2. For all database scenarios, we employ optimization solvers and DNN/RL agents to generate the ground-truth solutions and model parameters.

\vspace{0.3em}
\noindent
\textbf{Baselines.} We benchmark GridRAG against three categories of methods: 1) Cross-Scenario Approaches: Knowledge Graph (KG)~\cite{liu2022varying,gao2023data} and Meta-Learning (Meta-L)~\cite{luo2023generalizable,ding2024load}. 2) DNN and RL. According to conventional practices, DNN is applied to Task B, and RL is applied to Tasks A and C. 3) Model-based, which directly uses solvers to solve.

\begin{table*}[t]
\centering
\caption{Main results. Best results are in bold, and the second best are underlined.}
\label{tab:main results}
\renewcommand{\arraystretch}{1.2}  
\setlength{\tabcolsep}{5pt} 
\resizebox{\textwidth}{!}{
\begin{tabular}{c|c||cc|cc|cc||cc|cc||cc|cc|cc}
\toprule
\multicolumn{2}{c||}{\multirow{3}{*}{Methods}} & \multicolumn{10}{c||}{  \textit{\textbf{Test Group 1: Retrieved Similarity}}} & \multicolumn{6}{c}{ \textit{\textbf{Test Group 2: Zero-Shot}}} \\
\cline{3-18}
\multicolumn{2}{c||}{} & \multicolumn{2}{c|}{GridRAG*} & \multicolumn{2}{c|}{KG} & \multicolumn{2}{c||}{Meta-L} & \multicolumn{2}{c|}{DNN/RL} & \multicolumn{2}{c||}{model-based} & \multicolumn{2}{c|}{GridRAG} & \multicolumn{2}{c|}{DNN/RL} & \multicolumn{2}{c}{model-based}\\
\cline{3-18}
\multicolumn{2}{c||}{} & time/s & obj. gap & time/s & obj. gap & time/s & obj. gap & time/s & obj. gap & time/s & obj. gap & time/s & obj. gap & time/s & obj. gap & time/s & obj. gap \\
\midrule
\multirow{5}{*}{\rotatebox{90}{Task A}} 
& 13 & \textbf{1.23} & \textbf{1.07\%} & \underline{1.55} & 1.35\% & 16.87 & \underline{1.12\%} & 85.45 & 1.56\% & 10.32 & / & \textbf{21.37} & / & 89.28 & 3.37\% & 11.57 & / \\
& 33 & \textbf{1.51} & \underline{0.68\%} & \underline{1.70} & \textbf{0.61\%} & 16.75 & 0.68\% & 193.44 & 0.71\% & 917.85 & / & \textbf{134.19} & / & 187.60 & 1.80\% & 889.36 & / \\
& 69 & \textbf{2.35} & \textbf{0.54\%} & \underline{2.70} & 0.61\% & 18.88 & \underline{0.60\%} & 214.55 & 0.70\% & 975.60 & / & \textbf{122.72} & / & 225.87 & 2.07\% & 962.83 & / \\
& 123 & \textbf{3.98} & \underline{0.70\%} & \underline{5.01} & \textbf{0.64\%} & 21.29 & 0.77\% & 237.53 & 0.82\% & 2882.19 & / & \textbf{219.68} & / & 235.08 & 3.54\% & 2944.54 & / \\
\cline{2-18}
& Avg & \textbf{2.27} & \textbf{0.75\%} & \underline{2.74} & 0.80\% & 18.45 & 0.79\% & 182.74 & 0.95\% & 1196.49 & / & \textbf{124.49} & / & 184.46 & 2.70\% & 1202.08 & / \\
\midrule
\multirow{5}{*}{\rotatebox{90}{Task B}} 
& 13 & \textbf{1.25} & \textbf{2.00\%} & \underline{1.58} & \underline{2.00\%} & 17.44 & 2.24\% & 15.37 & 4.21\% & 2.59 & / & 13.87 & / & 15.58 & 15.95\% & \textbf{2.55} & / \\
& 33 & \textbf{1.49} & \underline{1.66\%} & \underline{1.70} & 1.78\% & 17.84 & \textbf{1.65\%} & 17.64 & 2.79\% & 5.68 & / & 14.04 & / & 17.33 & 13.71\% & \textbf{5.55} & / \\
& 69 & \textbf{2.46} & \underline{3.01\%} & \underline{2.82} & 3.09\% & 18.60 & \underline{3.09\%} & 15.74 & 3.26\% & 5.72 & / & 15.15 & / & 15.79 & 11.21\% & \textbf{5.70} & / \\
& 123 & \textbf{4.33} & \textbf{1.98\%} & \underline{5.11} & \underline{2.12\%} & 19.10 & 2.88\% & 16.68 & 4.57\% & 9.78 & / & 18.25 & / & 17.05 & 15.31\% & \textbf{9.74} & / \\
\cline{2-18}
& Avg & \textbf{2.38} & \textbf{2.16\%} & \underline{2.80} & \underline{2.25\%} & 18.25 & 2.47\% & 16.36 & 3.71\% & 5.94 & / & 15.33 & / & 16.44 & 14.04\% & \textbf{5.89} & / \\
\midrule
\multirow{5}{*}{\rotatebox{90}{Task C}} 
& 13 & \textbf{1.47} & \textbf{3.99\%} & \underline{1.68} & \underline{4.99\%} & 22.42 & 6.00\% & 373.14 & 6.53\% & 3.32 & / & 15.38 & / & 369.52 & 11.52\% & \textbf{3.43} & / \\
& 33 & \textbf{1.83} & \textbf{2.11\%} & \underline{1.92} & 2.18\% & 23.18 & \underline{2.15\%} & 545.51 & 2.38\% & 180.49 & / & \textbf{36.32} & / & 522.37 & 11.89\% & 185.60 & / \\
& 69 & \underline{3.01} & \textbf{1.75\%} & \textbf{2.84} & \underline{3.00\%} & 24.75 & 3.48\% & 240.95 & 3.87\% & 209.96 & / & \textbf{45.54} & / & 248.55 & 17.82\% & 218.71 & / \\
& 123 & \textbf{5.75} & \textbf{5.09\%} & \underline{5.93} & \underline{6.63\%} & 26.56 & 8.77\% & 424.22 & 9.36\% & 3251.11 & / & \textbf{271.92} & / & 440.91 & 25.39\% & 3209.82 & / \\
\cline{2-18}
& Avg & \textbf{3.02} & \textbf{3.24\%} & \underline{3.09} & \underline{4.20\%} & 24.23 & 5.10\% & 395.96 & 5.54\% & 911.22 & / & \textbf{92.29} & / & 395.34 & 16.66\% & 904.39 & / \\
\bottomrule
\end{tabular}
}
\vspace{-0.5em}
\end{table*}

\noindent
\textit{Note:} To isolate the performance gains introduced by the RAG paradigm, all end-effectors (DNN, RL, Solvers) utilize standard base architectures without additional ad-hoc acceleration heuristics. GridRAG acts as a plug-and-play framework compatible with any downstream model.

\vspace{0.3em}
\noindent
\textbf{Evaluation Metrics.} We use three key metrics: Time Consumption $\downarrow$, Objective Gap $\downarrow$, and Constraint Violation Rate $\downarrow$. Time consumption refers to the time taken to complete the task, so this varies in different schemes: Encoding + Retrieval + Refinement + Solving for GridRAG; Encoding + Searching + Direct Inference / Fine-tuning for KG; Encoding + Adaptation + Matching + Inference for Meta-L; Retraining + Inference for DNN/RL; Solving for model-based.

To ensure a fair comparison in Test Group 1, we calibrate the similarity thresholds for the KG and Meta-L baselines to prevent triggering retraining steps. Furthermore, we define a variant, GridRAG*, which disables diffusion refinement in favor of standard fine-tuning (same as KG and Meta-L). In Test Group 2, we omit KG and Meta-L. As stated in their original papers, they intrinsically degenerate to standard DNN/RL when threshold conditions fail in severe OOD scenarios.

\begin{figure}[!t]
\centering
\includegraphics[width=\columnwidth]{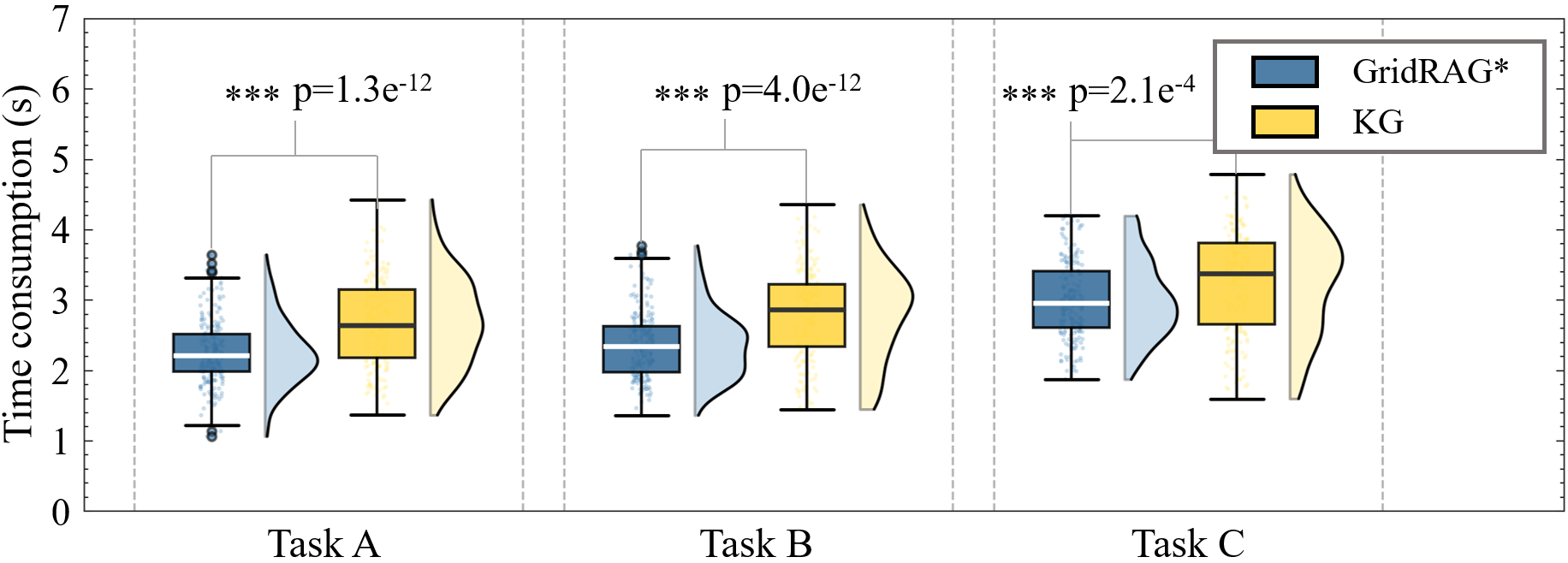}
\caption{Time consumption of GridRAG* and KG}
\label{fig:x}
\vspace{-0.5em}
\end{figure}

\vspace{0.3em}
\noindent
\textbf{Implementation details.} Database optimization utilizes the Gurobi solver with a predefined gap of $1e^{-4}$. The encoder in Grid2Vec is pre-trained. The diffusion module implement a cosine noise schedule over a maximum of 500 timesteps. The denoising network employs a Transformer structure with 3 encoder layers, 2 decoder layers, a hidden dimension of ${{d}_{\text{model}}}=64$, and 4 attention heads. During SDEdit-based inference, the noise level is controlled by retrieval confidence to the range [0.2,0.6]. We optimize the training parameters of all baselines while keeping their model structure unchanged. Sepecificlly, for the KG baseline, we conduct knowledge graph encoded with network structure, training hyperparameters and DER scenarios~\cite{liu2022varying,gao2023data}. For Meta-L baseline, we strictly follow the model designs in~\cite{luo2023generalizable,ding2024load}, where a lightweight 1D-CNN is used as the encoder and a RelationNet-style nonlinear comparator as the relation module. All experiments are conducted on Intel Xeon Gold 6254 processor and NVIDIA A800 GPU.

\vspace{-0.4em}
\subsection{Main Results}

Table \uppercase\expandafter{\romannumeral1} presents the performance metrics of all evaluated models under two sets of test conditions. Since the commercial solver is considered to provide the optimal ground truth, its objective gaps are denoted by ``/''.

\textbf{1) Performance on Test Group 1}.
In scenario-specific baselines, model-based approaches typically incur substantial computational overhead; notably, tasks A and C, which involve numerous discrete variables, require several hours to solve. DNN/RL’s accuracy degrades significantly during cross-scenario inference, exhibiting an objective gap of nearly 10\% on Task C. And retraining these models incurs prohibitive time and computational costs.

In cross-scenario baselines, both KG and Meta-L avoid the need for retraining, accomplishing the tasks relying solely on online inference. However, Meta-L yields the highest objective gap among all cross-scenario methods, indicating its heavy reliance on in-distribution conditions.

The proposed GridRAG achieves the fastest solution times and the highest accuracy across most tasks. The average time consumption across all tasks is kept under 3s, which is an acceleration of over 300 times compared to solvers and over 100 times compared to retraining DNN/RL. This underscores the time-efficiency advantage of the RAG framework for online cross-scenario inference. It also demonstrates a statistically significant improvement over other cross-scenario methods (p$<$0.05), as illustrated in Fig. 4. This is primarily attributed to our two-stage hybrid retrieval, which rapidly filters similar scenarios from massive candidate pools.

\textbf{2) Performance on Test Group 2}.
Under group 2, where new scenarios exhibit severe distribution shifts, the inference accuracy of DNN/RL becomes unacceptable, reaching a 25\% on Task C. By contrast, GridRAG leverages diffusion refinement to generate a high-quality initial solution for warm-start, which subsequently reduces the solver's computational time by more than 10 times.

\begin{table}[t]
\centering
\caption{Violation Rate Comparison}
\label{tab:violation}
\setlength{\tabcolsep}{3pt}
\resizebox{\columnwidth}{!}{
\begin{tabular}{cc||cccc||cc}
\toprule
\multicolumn{2}{c||}{\multirow{2}{*}{Viol.\ Rate $\downarrow$}}
  & \multicolumn{4}{c||}{Test Group 1}
  & \multicolumn{2}{c}{Test Group 2} \\
\cmidrule(lr){3-6} \cmidrule(lr){7-8}
\multicolumn{2}{c||}{}
  & GridRAG* & KG & Meta-L. & DNN\,/\,RL & GridRAG & DNN\,/\,RL \\
\midrule
\multirow{4}{*}{\rotatebox[origin=c]{90}{Task A}}
  & 13  & 0.00\% & 0.00\% & 0.00\% & 0.00\% & 0.00\% & 0.00\% \\
  & 33  & 0.00\% & 0.00\% & 0.00\% & 0.01\% & 0.00\% & 0.04\% \\
  & 69  & 0.02\% & 0.02\% & 0.04\% & 0.18\% & 0.00\% & 0.69\% \\
  & 123 & 0.16\% & 0.16\% & 0.16\% & 0.20\% & 0.00\% & 0.44\% \\
\midrule
\multirow{4}{*}{\rotatebox[origin=c]{90}{Task B}}
  & 13  & 3.32\% & 4.30\% & 3.08\% & 6.25\% & 0.00\% & 8.99\% \\
  & 33  & 2.04\% & 3.71\% & 2.10\% & 4.08\% & 0.00\% & 4.26\% \\
  & 69  & 1.07\% & 1.77\% & 1.88\% & 3.78\% & 0.00\% & 6.38\% \\
  & 123 & 1.55\% & 1.63\% & 0.76\% & 1.95\% & 0.00\% & 2.86\% \\
\midrule
\multirow{4}{*}{\rotatebox[origin=c]{90}{Task C}}
  & 13  & 0.00\% & 0.00\% & 0.00\% & 0.00\% & 0.00\% & 0.00\% \\
  & 33  & 0.02\% & 0.02\% & 0.02\% & 0.02\% & 0.00\% & 0.02\% \\
  & 69  & 0.00\% & 0.00\% & 0.00\% & 0.00\% & 0.00\% & 0.03\% \\
  & 123 & 0.26\% & 0.40\% & 0.40\% & 0.42\% & 0.00\% & 1.40\% \\
\bottomrule
\end{tabular}
}
\vspace{-0.5em}
\end{table}

\begin{figure*}[!t]
\centering
\includegraphics[width=\textwidth]{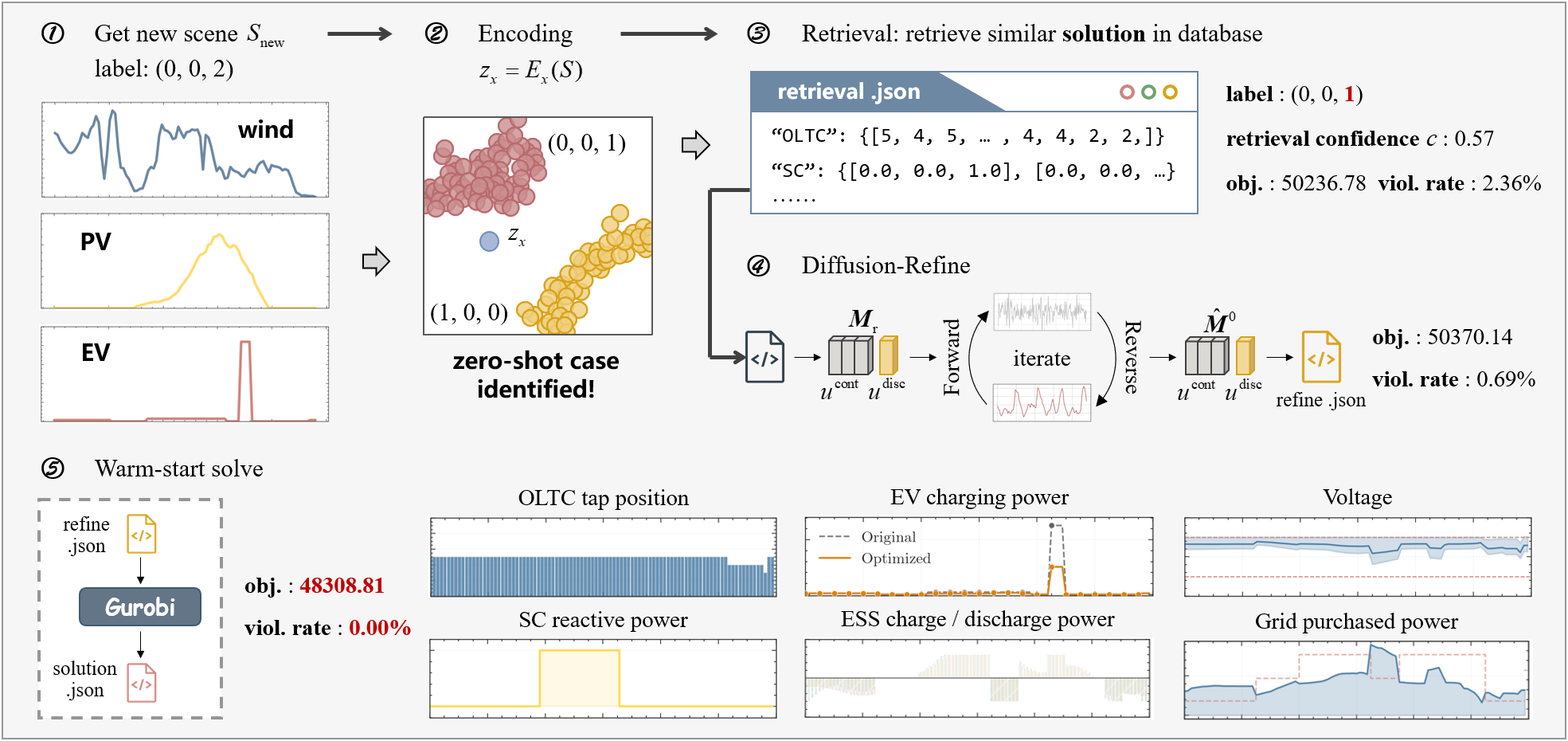}
\caption{GridRAG's workflow: case study on Task C, IEEE-33 bus system}
\label{fig:x}
\vspace{-0.5em}
\end{figure*}

\textbf{3) Constraint Violation Analysis}.
Table \uppercase\expandafter{\romannumeral2} reports the constraint violation rates. GridRAG and other cross-scenario methods demonstrate comparable violation rates, which are strictly lower than those of direct DNN/RL inference. Notably, while constraint violations inherently increase under zero-shot conditions, GridRAG guarantees ultimate physical feasibility thanks to the final refinement step executed by the exact solver.

\vspace{-0.4em}
\subsection{Case Study}

In this section, we present a comprehensive and representative case study to demonstrate: 1) the step-by-step workflow of GridRAG; 2) an interpretable analysis of the generated results.

As shown in Fig. 5, the process begins with the input of a new, previously unseen scenario labeled as (0,0,2), a combination excluded from the database. Specifically, scenarios with an EV profile label of 2 denotes centralized, ultra-fast charging behaviors in the late afternoon~\cite{gaete2021open}. This case effectively mimics a severe distribution shift, which serves as a strong test for zero-shot generalization capability.

After embedding the input scenario using Grid2Vec, we visualize its position in the learned latent space via t-SNE. The red and yellow clusters represent known label groups in the database, while the blue point (our test scenario) lies clearly outside all existing clusters, confirming that this is a genuine zero-shot case.

The Hybrid Retrieval module then retrieves the most similar historical scenarios and their corresponding solutions, and outputs a retrieval confidence score (as defined in Eq. (4)) to control the strength of downstream refinement.

In the next step, the Diffusion-Refinement module performs a forward diffusion process whose maximum timestep is adaptively determined by the retrieval confidence (see Eq. (13)). The denoising network remains frozen during inference.

The final optimized solution achieves a total operating cost of 48,308.81, substantially lower than the cost of the raw retrieved solution (50,236.78), while strictly satisfying all physical constraints. To interpret these results physically, we plot the trajectories of key control and grid variables in the bottom of Fig. 5:

\begin{itemize}[leftmargin=9pt]
  \item The OLTC tap position remains at or near the highest tap for most of the day; SC provide more reactive compensation around noon, helping reduce active power losses and alleviate voltage violations.
  \item Because the database contains no prior knowledge of the centralized ultra-fast EV charging behavior, the raw retrieved solution from label (0,0,1) incurs a higher cost and violates constraints by up to 2.36\%. The refined solution intelligently curtails peak EV load, forming what we consider a highly near-feasible solution for the unseen label (0,0,2).
  \item The node voltages are strictly maintained within the safe 0.95-1.05 p.u. range; The model strategically increases grid power purchases and charges ESS during off-peak afternoon and night periods when electricity prices (red line) are lower.
\end{itemize}

Fig. 6 compares the end-to-end computational time of different approaches on this case. Retraining a task-specific RL model requires approximately 500 s, while directly solving with Gurobi takes around 200 s. In contrast, the majority of GridRAG’s time is spent on the diffusion inference, and the refined warm-start solution enables the solver to converge in under 10 s, resulting in a dramatic reduction in total runtime.

\begin{figure}[!t]
\centering
\includegraphics[width=\columnwidth]{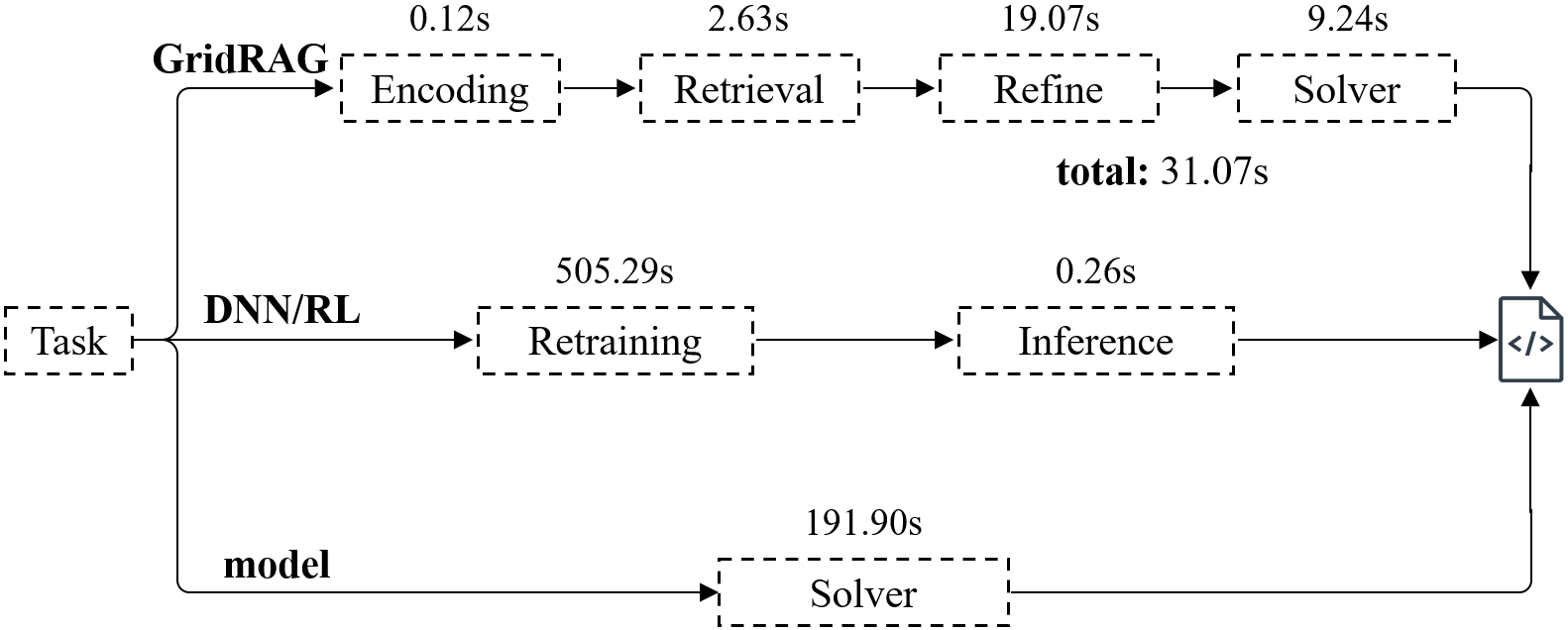}
\caption{Time consumption in each stage of different methods}
\label{fig:x}
\vspace{-0.3em}
\end{figure}

\subsection{Ablation and Discussion}

\noindent
\textbf{Ablation study.} Table III details the ablation study of GridRAG's components. Removing Grid2Vec increases the objective gap by 2\% despite similar time consumption, highlighting the necessity of multi-modal semantic consistency over raw numerical proximity scenario retrieval. Replacing Hybrid Retrieval with global search significantly increases computational overhead. Furthermore, omitting the diffusion-refinement module causes raw retrieved solutions to perform poorly as warm-starts. Completely removing the warm-start solver phase reduces execution time but severely degrades solution quality.

\begin{table}[h]
\centering
\caption{Ablation Study on Model Design}
\label{tab:ablation}
\renewcommand{\arraystretch}{1.1}  
\resizebox{\columnwidth}{!}{
\begin{tabular}{l|cc|cc|cc}
\toprule
& \multicolumn{2}{c|}{Task A} & \multicolumn{2}{c|}{Task B} & \multicolumn{2}{c}{Task C} \\
\cline{2-7}

& tiem/s & obj. gap & tiem/s & obj. gap & tiem/s & obj. gap \\
\midrule

GridRAG* & 2.27 & 0.75\% & 2.38 & 2.16\% & 3.02 & 3.24\% \\
w/o Grid2Vec & 2.21 & 2.89\% & 2.28 & 3.91\% & 2.90 & 5.52\% \\
w/o Hyb. Retr. & 7.55 & 0.76\% & 7.67 & 2.16\% & 8.03 & 3.27\% \\
\midrule
GridRAG & 124.49 & / & 15.33 & / & 92.29 & / \\
w/o Diff. & 941.26 & / & 5.57 & / & 782.58 & / \\
w/o warm-start & 36.5 & 3.79\% & 8.82 & 13.11\% & 29.84 & 22.03\% \\
\bottomrule
\end{tabular}
}
\vspace{-0.3em}
\end{table}

\noindent
\textbf{Evaluation of Multi-Modal Retrieval.} To validate our multi-modal insight, we evaluate retrieval performance in the joint space versus a scene-only space. We disable the downstream fine-tuning and refinement modules to directly assess the raw quality of the retrieved solutions. We define database solution closest to the ground truth as \textit{oracle}. The results in test group 1 are shown in Table IV. Our joint embedding with hybrid retrieval closely approximates this upper bound, whereas scene-only retrieval yields a larger objective gap. Furthermore, Fig. 7 visualizes zero-shot scenario distributions. Clearly, the new scenario is distant from known clusters in the scene space. However, in the joint space, it is positioned very close to the manifold of the red cluster. This demonstrates that scenario distribution shifts do not strictly imply solution heterogeneity, firmly validating our multi-modal design.

\begin{table}[t]
\centering
\caption{Multi-Modal Evaluation on Test Group 1}
\label{tab:ablation}
\setlength{\tabcolsep}{4pt}
\renewcommand{\arraystretch}{1.1}  
\resizebox{\columnwidth}{!}{
\begin{tabular}{l|cc|cc|cc}
\toprule
& \multicolumn{2}{c|}{Task A} & \multicolumn{2}{c|}{Task B} & \multicolumn{2}{c}{Task C} \\
\cline{2-7}

& viol. rate & obj. gap & viol. rate & obj. gap & viol. rate & obj. gap \\
\midrule

scene-only & 0.50\% & 1.66\% & 6.65\% & 4.31\% & 1.28\% & 8.89\% \\
hybrid & 0.39\% & 1.06\% & 6.04\% & 3.97\% & 1.08\% & 7.03\% \\
oracle & 0.14\% & 0.89\% & 4.88\% & 3.10\% & 0.44\% & 4.72\% \\
\bottomrule
\end{tabular}
}
\vspace{-0.5em}
\end{table}

\begin{figure}[h]
\centering
\includegraphics[width=0.9\columnwidth]{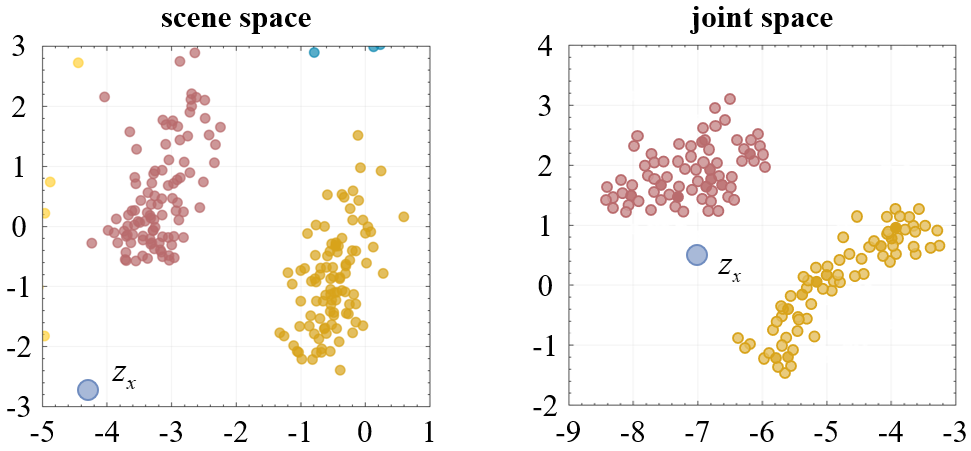}
\caption{Multi-modal validation: zero-shot scenario distribution in scene space and joint space}
\label{fig:x}
\vspace{-0.3em}
\end{figure}

More ablation and hyperparameter sensitivity analysis are provided in Appendix B.

\section{Conclusion}
In this study, we introduce GridRAG, the first retrieval-augmented framework designed specifically for the optimal operation of distribution networks. GridRAG seamlessly integrates multi-modal representation learning, a two-stage fast retrieval mechanism, and conditional diffusion-based refinement. Comprehensive evaluations across various operational tasks and standard IEEE 13, 33, 69, and 123-bus test systems demonstrate that GridRAG drastically reduces online computational time. Furthermore, it achieves robust zero-shot transfer on severe out-of-distribution scenarios without the need for computationally expensive model retraining. The complete source code for this project has been made publicly available at https://anonymous.4open.science/r/GridRAG-A328.


%



\section*{Acknowledgment}


This work was supported by XXX.

\ifCLASSOPTIONcaptionsoff
  \newpage
\fi



%



\bibliographystyle{IEEEtran}




\vspace{-0.3em}
\section*{Appendix}

\subsection{Optimization Problem Formulations}

\renewcommand{\thefigure}{A\arabic{figure}}
\setcounter{figure}{0}
\renewcommand{\theequation}{A\arabic{equation}}
\setcounter{equation}{0}
\renewcommand{\thetable}{A\arabic{table}}
\setcounter{table}{0}

\textit{Due to space limitations, a complete variable nomenclature table  
will be provided in the second-round revision.}

\noindent
\textbf{1) Network topology and DER placement}

We evaluate GridRAG on four IEEE radial test feeders and set DER location following~\cite{chen2025large,chen2025robust} as shown in Table A1:

\begin{table}[h] 
\vspace{-0.3em}
    \centering 
    \caption{DER placement across test systems}
    \label{tab:dataset} 

    \setlength{\tabcolsep}{5pt} 
    \resizebox{\columnwidth}{!}{ 
    \begin{tabular}{ c| c c c c}
        \toprule
        \textbf{Device} & \textbf{IEEE-13} & \textbf{IEEE-33} & \textbf{IEEE-69} & \textbf{IEEE-123} \\ 
        \midrule
        PV & 634, 580, 684 & 18, 33 & 27, 50, 62 & 32, 50, 83, 88, 110 \\
        WT & / & 22, 25 & 18, 65 & 41, 57, 71 \\
        ESS & 671 & 10, 23, 31 & 12, 50 & 50, 88 \\
        EVs & 692 & 15, 21, 29 & 28, 51, 66 & 48, 65, 76, 96, 114 \\
        SC & 611, 675 & 16, 20, 31 & 12, 50, 62 & 65, 83, 88 \\
        SVC & 671, 675 & 11, 29 & 18, 51 & 50, 71, 110 \\
        \midrule
        Tie lines & 0 & 3 & 5 & 3 \\
        \bottomrule
    \end{tabular}
    }
    \vspace{-0.6em}
\end{table}

\noindent
\textbf{2) Task A — Volt-VAR Control}

Task A addresses the distribution VVC problem. The objective minimizes network loss with a voltage-violation penalty:

\vspace{-0.6em}
\begin{equation}\label{eq:a1}
\min\;\sum_{t\in\mathcal{T}}\sum_{(i,j)\in\mathcal{L}} r_{ij}\,\ell_{ij,t} \;+\; \mu\sum_{t,j}\bigl(\xi_{j,t}^{+}+\xi_{j,t}^{-}\bigr)
\end{equation}

\noindent
subject to (main constraints):

\vspace{-0.6em}
\begin{equation}\label{eq:a2}
\sum\nolimits_{t=0}^{T-2}\phi_t\le\overline{A}^{\text{oltc}},\qquad \sum\nolimits_{t=0}^{T-2}\delta_{t,b}^{\text{sc}}\le\overline{A}^{\text{sc}}
\end{equation}

\vspace{-0.6em}
\begin{equation}\label{eq:a3}
Q_{t,b}^{\text{sc}} = n_{t,b}^{\text{sc}} \cdot q^{\text{step}}, \qquad
0 \le n_{t,b}^{\text{sc}} \le \bar{n}^{\text{sc}}
\end{equation}

\vspace{-0.6em}
\begin{equation}\label{eq:a4}
\bigl( Q_{t,i}^{\text{pv}} \bigr)^2
  \le \bigl( \bar{S}_i^{\text{pv}} \bigr)^2
     - \bigl( P_{t,i}^{\text{pv}} \bigr)^2
\end{equation}

Constraints (A2) limit daily switching actions of the OLTC and each SC bank. Eq. (A3) links SC reactive output to its discrete stage. Eq. (A4) enforces the PV inverter apparent-power limit (analogous for WT).

\noindent
\textbf{3) Task B — Economic Dispatch}

Task B focuses on cost-driven economic operation, co-optimizing grid power purchase, ESS charge-discharge, PV curtailment, and network reconfiguration via tie-switch operations. The daily operating cost is:

\vspace{-0.8em}
\begin{equation}\label{eq:a5}
\resizebox{1.05\hsize}{!}{
$\min\;
  \sum_{t} \Bigl[
    C_t^{\text{e}}\, P_t^{\text{g}}
    + C^{\text{s}} \!\sum_{m} \bigl( P_{t,m}^{\text{ch}} + P_{t,m}^{\text{dis}} \bigr)
    + C^{\text{cut}} \!\sum_{i} P_{t,i}^{\text{cut}}
  \Bigr] \Delta t
  + C^{\text{w}} \!\sum_{t,m} \delta_{t,m}^{\text{sw}}$
  }
\end{equation}

\noindent
subject to (main constraints):

\vspace{-0.8em}
\begin{equation}\label{eq:a6}
E_{t,m} = E_{t-1,m}
  + \eta^{\text{ch}}\, P_{t,m}^{\text{ch}}\, \Delta t
  - {P_{t,m}^{\text{dis}}}{\eta^{\text{dis}}}\, \Delta t
\end{equation}

\vspace{-0.8em}
\begin{equation}\label{eq:a7}
P_{t,m}^{\text{ch}} \le \overline{P}_m^{\text{ch}}\,(1 - y_{t,m}), \qquad
P_{t,m}^{\text{dis}} \le \overline{P}_m^{\text{dis}}\, y_{t,m}
\end{equation}

\vspace{-0.8em}
\begin{equation}\label{eq:a8}
\bigl| P_{t,m}^{\text{tie}} \bigr| \le \Gamma\,\tau_{t,m}\, \
\bigl| Q_{t,m}^{\text{tie}} \bigr| \le \Gamma\,\tau_{t,m}\, \
\ell_{t,m}^{\text{tie}} \le \Gamma\,\tau_{t,m}
\end{equation}

Eqs. (A6)-(A7) govern ESS energy dynamics and charge–discharge mutual exclusion via mode binary $y_{t,k}$. Terminal SOC is constrained within $\pm 10\%$ of its initial value. Eq. (A8) enforces Big-M conditions on tie-line power flow.

\noindent
\textbf{4) Task C — Active/Reactive Coordinated Optimization}

Task C integrates all devices from Tasks A and B and additionally incorporating EV charging stations with demand-side flexibility. The objective extends (A5) with voltage and EV penalties:

\vspace{-0.8em}
\begin{equation}\label{eq:a9}
\min\;
  \text{(A5)}
  + \mu \!\sum \bigl( \xi_{j,t}^{+} + \xi_{j,t}^{-} \bigr)
  + \Phi^{\text{ev}}
\end{equation}

\noindent
with the following EV-specific constraints:

\vspace{-0.8em}
\begin{equation}\label{eq:a10}
P_{t,m}^{\text{ev}} = \hat{P}_{t,m}^{\text{ev}}\,(1 - \rho_{t,m})\, \, \
\rho_{t,m} = \rho_{t,m}^{(1)} + \rho_{t,m}^{(2)} + \rho_{t,m}^{(3)}
\end{equation}

\vspace{-0.8em}
\begin{equation}\label{eq:a11}
\Phi_{\text{tier}}^{\text{ev}}
  = \sum \hat{P}_{t,m}^{\text{ev}} \bigl(
      C_1^{\text{ev}}\,\rho_{t,m}^{(1)}
    + C_2^{\text{ev}}\,\rho_{t,m}^{(2)}
    + C_3^{\text{ev}}\,\rho_{t,m}^{(3)}
  \bigr) \Delta t
\end{equation}

\vspace{-0.8em}
\begin{equation}\label{eq:a12}
E_{t,m}^{\text{ev}} = E_{t-1,m}^{\text{ev}}
  + \eta^{\text{ev}}\, P_{t,m}^{\text{ev}}\, \Delta t,\, \, \
\zeta_m \ge E_m^{\text{target}} - E_{T,m}^{\text{ev}}
\end{equation}

\vspace{-0.8em}
\begin{equation}\label{eq:a13}
z_{t,m}^{\text{int}} \ge w_{t-1,m} - w_{t,m}, \qquad \forall\, t \ge 2
\end{equation}

\vspace{-0.8em}
\begin{equation}\label{eq:a14}
\Phi^{\text{ev}}
  = \Phi_{\text{tier}}^{\text{ev}}
  + C^{\text{short}} \!\sum \zeta_m
  + C^{\text{int}} \!\sum z_{t,m}^{\text{int}}
\end{equation}

Eq. (A10) defines controllable EV charging power, where $\rho_{t,m}$ is the curtailment ratio with increasing marginal penalties $C_1^{ev}<C_2^{ev}<C_3^{ev}$ in Eq. (A11). Eq. (A12) tracks accumulated EV energy and defines shortage slack $\zeta_m$. Eq. (A13) detects charging interruptions via binary $z_{t,m}^{\text{int}}$. Additionally, Task C inherits all device constraints from Tasks A and B.

\vspace{-0.5em}
\subsection{More Ablation and Hyperparameter Sensitivity Analysis}

\renewcommand{\thefigure}{B\arabic{figure}}
\setcounter{figure}{0}
\renewcommand{\theequation}{B\arabic{equation}}
\setcounter{equation}{0}
\renewcommand{\thetable}{B\arabic{table}}
\setcounter{table}{0}

\noindent
\textbf{To what extent does Diffusion-Refine contribute to warm-start?} Fig. B1 presents confusion matrices indicating valid warm-starts (`1' for solver acceptance/acceleration, `0' for rejection) before and after diffusion-refinement in zero-shot settings. Directly retrieved solutions form effective warm-starts in under 10\% of cases. In contrast, refined solutions succeed 95\% of the time, proving the efficacy of our SDEdit-style diffusion. Table B1 further explains this mechanism via context-FID~\cite{chen2025chime} and correlation matrix distances between solution distributions and the ground-truth warm-start basin of attraction. For both metrics, lower values indicate higher similarity. The refined distributions exhibit remarkably high similarity to the optimal basin, providing clear interpretability for GridRAG’s robust zero-shot performance.

\begin{figure}[h]
\centering
\includegraphics[width=1.0\columnwidth]{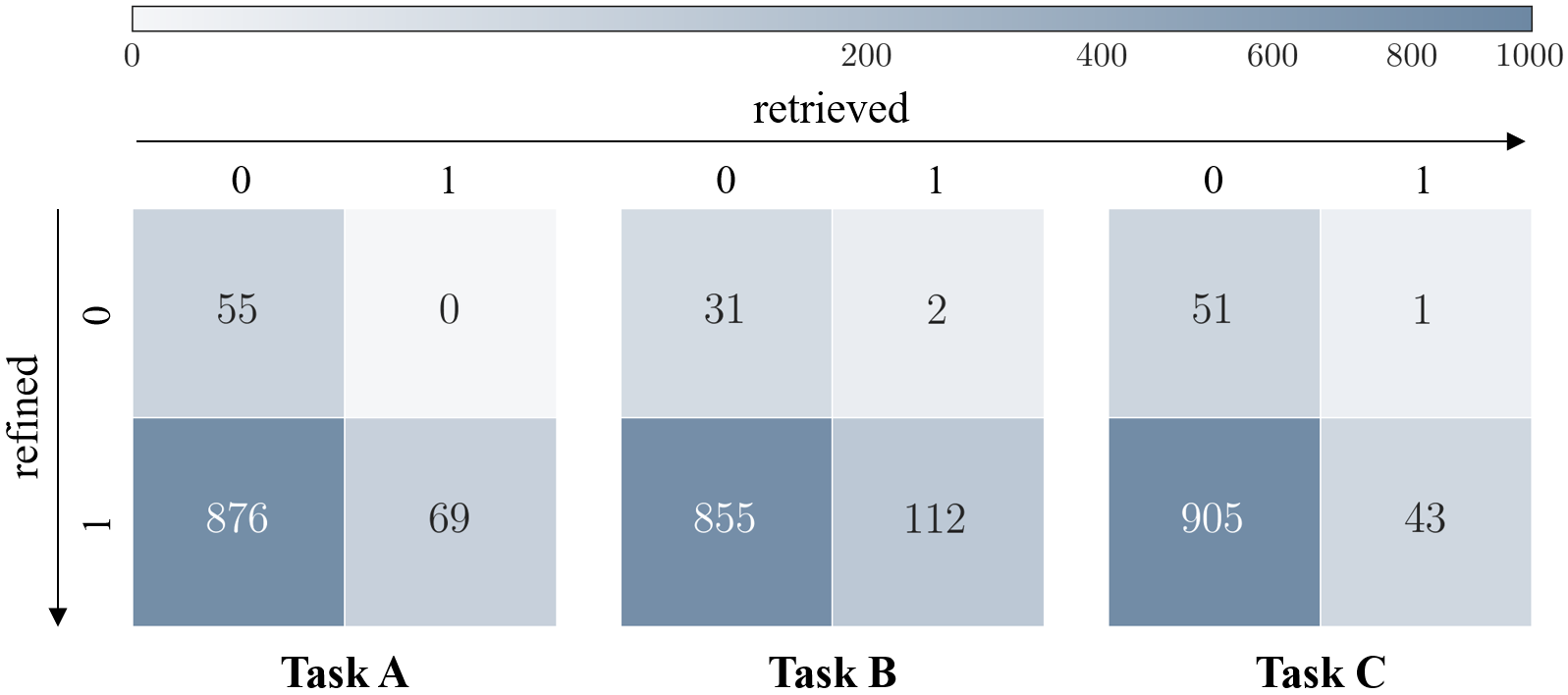}
\caption{Confusion matrix: number of valid warm-starts}
\label{fig:x}
\vspace{-0.3em}
\end{figure}

\begin{table}[h]
\centering
\caption{Similarity matrix between solution distributions and warm-start basin}
\label{tab:ablation}
\setlength{\tabcolsep}{3pt}
\renewcommand{\arraystretch}{1.2}  
\resizebox{\columnwidth}{!}{
\begin{tabular}{l|cc|cc|cc}
\toprule
& \multicolumn{2}{c|}{Task A} & \multicolumn{2}{c|}{Task B} & \multicolumn{2}{c}{Task C} \\
\cline{2-7}

& FID & Corr. & FID & Corr. & FID & Corr. \\
\midrule

retrieved & 0.205$\pm$.017 & 0.186$\pm$.018 & 0.187$\pm$.023 & 0.160$\pm$.019 & 0.300$\pm$.023 & 0.220$\pm$.011 \\
refined & 0.028$\pm$.002 & 0.017$\pm$.001 & 0.022$\pm$.002 & 0.014$\pm$.000 & 0.027$\pm$.003 & 0.015$\pm$.002 \\

\bottomrule
\end{tabular}
}
\vspace{-0.3em}
\end{table}

\noindent
\textit{Note:} The results reported in Table I (Main Results) represent the average execution times including those where a warm-start failed to form. Thus, successful warm-starts achieve even shorter solving times than listed. The ground-truth basin of attraction is empirically bounded by the maximum range within the same test cluster that triggers an effective warm-start. The correlation matrix distance is defined as:

\vspace{-0.8em}
\begin{equation}\label{eq:b1}
\frac{1}{10}\sum\limits_{i,j}{\frac{\operatorname{cov}_{i,j}^{\text{warm}}}{\sqrt{\operatorname{cov}_{i,i}^{\text{warm}}\cdot \operatorname{cov}_{j,j}^{\text{warm}}}}-\frac{\operatorname{cov}_{i,j}^{\text{re}}}{\sqrt{\operatorname{cov}_{i,i}^{\text{re}}\cdot \operatorname{cov}_{j,j}^{\text{re}}}}}
\end{equation}

\noindent
\textbf{Analysis of k.} In Fast Recall the performance is relatively robust to $K_0$; however, excessively small values ($<$20) degrade downstream accuracy, as scene-only recall may entirely miss the \textit{oracle}. For most tasks, setting $k=2$ yields optimal results on test group 1, achieving an 8\% accuracy improvement over $k=1$. For test group 2, $k=1$ is preferred.

\vspace{0.2em}
\noindent
\textbf{Single vs. Unified Database.} We explored a unified database for all topologies by integrating a GCN-based encoder~\cite{chen2025egbad} into Grid2Vec to capture network structures. Results indicate that while a unified database maintains solution accuracy, it increases computational overhead by 6.3\%, 6.7\%, and 7.9\% for Tasks A, B, and C, respectively. This latency stems from both the encoding and retrieval phases. Thus, maintaining topology-specific databases is more efficient.

\vspace{0.2em}
\noindent
\textbf{Practical Discussion.} GridRAG is a plug-and-play framework compatible with various downstream optimize strategies. The diffusion module can use other generative backbones such as flow-based models. While this study focuses on distribution networks heavily affected by DER uncertainty, the RAG strategy is readily extendable to IES, buildings, etc. Despite GridRAG’s superior zero-shot capabilities, we recommend continuous online database updates to further enhance long-term operational performance in real-world deployments.


%








\end{document}